\documentclass[12pt]{article}
\pdfoutput=1
\usepackage{latexsym, graphicx,cite} 
\usepackage{amsmath}
\usepackage{amssymb}
\numberwithin{equation}{section}

\usepackage[colorlinks=true, linkcolor=blue, bookmarks=true]{hyperref}

\makeatletter
\renewcommand\section{\@startsection {section}{1}{\z@}%
                                   {-3.5ex \@plus -1ex \@minus -.2ex}
                                   {2.3ex \@plus.2ex}%
                                   {\normalfont\large\bfseries}}
\renewcommand\subsection{\@startsection{subsection}{2}{\z@}%
                                     {-3.25ex\@plus -1ex \@minus -.2ex}%
                                     {1.5ex \@plus .2ex}%
                                     {\normalfont\bfseries}}
\makeatother

\newcommand{\beq}{\begin{equation}}
\newcommand{\eeq}{\end{equation}}
\newcommand{\bea}{\begin{eqnarray}}
\newcommand{\eea}{\end{eqnarray}}

\newcommand{\CC}{{\mathbb C}}
\newcommand{\C}{\CC}
\newcommand{\PP}{{\mathbb P}}

\begin{document}

\begin{titlepage}
\begin{flushright}
\phantom{arXiv:yymm.nnnn}
\end{flushright}
\vfill
\begin{center}
{\Large\bf A Hidden Symmetry of AdS Resonances}    \\
\vskip 15mm
{\large Oleg Evnin$^{a,b}$ and Chethan Krishnan$^c$}
\vskip 7mm
{\em $^a$ Department of Physics, Faculty of Science, Chulalongkorn University,\\
Thanon Phayathai, Pathumwan, Bangkok 10330, Thailand}
\vskip 3mm
{\em $^b$ Theoretische Natuurkunde, Vrije Universiteit Brussel and\\
The International Solvay Institutes\\ Pleinlaan 2, B-1050 Brussels, Belgium}
\vskip 3mm
{\em $^c$ Center for High Energy Physics, Indian Institute of Science, \\ 
  C V Raman Avenue, Bangalore 560012, India}

\vskip 3mm
{\small\noindent  {\tt oleg.evnin@gmail.com, chethan.krishnan@gmail.com}}
\vskip 10mm
\end{center}
\vfill

\begin{center}
{\bf ABSTRACT}\vspace{3mm}
\end{center}

Recent investigations have revealed powerful selection
rules for resonant energy transfer between modes of non-linear
perturbations in global anti-de Sitter (AdS) space-time. 
It is likely that these selection rules are due to the highly symmetric nature of the underlying AdS background, though the precise relation has remained unclear. In this 
article, we demonstrate that the equation satisfied by the scalar field mode functions in AdS$_{d+1}$ has 
a hidden $SU(d)$ symmetry, and explicitly specify the multiplets of this $SU(d)$ symmetry furnished by the mode functions. We also comment on the role this structure might
play in explaining the selection rules.

\vfill

\end{titlepage}


\section{Introduction and Overview}\label{introduction}\vspace{1.5mm}

\begin{flushright}
\emph{``The career of a young theoretical physicist\\ 
consists of treating the harmonic oscillator\\
 in ever-increasing levels of abstraction.''} \vspace{0.7mm}\\
(attributed to Sydney Coleman)\vspace{2mm}
\end{flushright}

\noindent 
The stability of AdS spacetimes \cite{BizonReview} against small amplitude (but non-linear) perturbations is an interesting question in its own right, as well as in light of the AdS/CFT correspondence. On the gravity side, a turbulent instability would be significant in the formation of black holes (possibly after repeated reflections of the perturbations from the boundary), and on the gauge theory side it would indicate thermalization in a strongly coupled quantum field theory at low energies. 

Investigations of this question have resulted in various surprises. In the pioneering article  \cite{Bizon1} it was found that weak amplitude Gaussian initial data in global AdS lead to black hole formation after multiple reflections from the boundary. The expectation there was that resonant energy transfer between the modes might be the reason behind the instability. Apparently non-collapsing initial data have been subsequently found in \cite{Maliborski:2013jca,Buchel:2013uba}. It was further claimed in \cite{Buchel1} that when only low-lying modes are turned on and the amplitude is small enough (parametrized by a dimensionless number $\epsilon$), the system does not collapse, at least for a very long time. The numerical results for the specific initial data considered in \cite{Buchel1} have been challenged in \cite{Bizon2}, but  it might still be possible that for sufficiently low-amplitude initial data, collapse can be delayed by much longer than $\sim 1/\epsilon^2$. Various other papers also suggest that the instability (if it exists) of low-lying low-amplitude modes might not be as virulent as it was originally thought \cite{CEV1, BKS1, Yang1, CEV2, Buchel2, Yang2, BKS2}.

Some understanding of the situation was gained in \cite{CEV1}, where using renormalization group resummation techniques, effective equations were derived, describing slow energy transfer between the modes. It was shown that the resonant energy transfer between modes in AdS is highly restricted, due to the fact that a majority of the possible growing (secular) terms in perturbation theory (signalling resonant energy transfer) do not in fact arise. In \cite{BKS1} it was argued that a probe quartic scalar field theory in AdS could be used as an instructive model for the full gravitational instability question -- among other things, it was noticed there  as well that there are restrictions on secular terms and resonant energy transfer. The description in \cite{BKS1} was couched in the language of the Two-Time Formalism (TTF) of \cite{Buchel1}, which is equivalent to the renormalization group (RG) resummation of \cite{CEV1} at relevant order. These results were further extended in \cite{Yang2} where the spherical symmetry assumption that was inherent in all previous discussions was dropped, and it was shown by direct manipulations with the mode functions that the scalar field still exhibits highly restrictive selection rules in its resonances. Another related observation is that the RG/TTF-resummed version of the theory possesses a number of conserved charges -- this structure was discovered for the scalar theory in \cite{BKS1} and for the full gravity theory in \cite{CEV2}. 

Our main motivation in this paper will be to gain a more concrete understanding of this vanishing of the various RG/TTF coefficients. We believe that the existence of resonance selection rules as well as the conservation laws is strongly indicative of some underlying symmetry principle. Note that explicit symmetries (and related conservation laws) can be identified in the RG/TTF effective equations \cite{BKS1, CEV2}, but this is not what concerns us here. Rather, we are looking for a symmetry of the underlying AdS background that leads to these conservation laws in the resummed effective theory. 

The fact that symmetries of the background can restrict the energy flow due to non-linearities among the perturbation modes is not unfamiliar in the context of non-linear system studies. For example, some symmetry-based selection rules have been discussed for vibrations of non-linear atomic lattices, in particular, in relation to the so-called `bushes' of modes -- these are subsets of modes only transferring energy to each other, but not to other modes, due to symmetry restrictions \cite{chechin}.

Our starting point in identifying the relevant symmetry will be that the frequencies of the scalar field mode functions in AdS background display a characteristic $2n+l$ degeneracy. Here, $n$ is a `radial quantum number' governing the radial dependence of the mode functions, and $l$ is the angular momentum governing their angular dependence. Typically, one expects degeneracies to arise when there are symmetries in the problem. Note however that this present degeneracy can {\em not} be understood in terms of spherical symmetry explicitly present in the equations. The spherical symmetry leads merely to the familiar fact that there is degeneracy in `azimuthal quantum numbers' for a given $l$. The fact that the frequencies only depend on $2n+l$, such that different choices of $n$ and $l$ produce the same frequency, clearly implies a bigger symmetry group.

In the rest of the paper, we will argue that this degeneracy can be explicitly understood in terms of a hidden symmetry in the mode function equation. We study the degenerate multiplets of AdS$_{d+1}$ mode functions, reveal that they form representations of (a rather non-obvious) $SU(d)$ group and explicitly specify the representation in which each mode function resides. The way we identify the hidden symmetry is by first relating the scalar wave equation in AdS to that in an Einstein static universe via a conformal transformation. Then we employ arguments similar in spirit to those associated to the $SO(4)$-enhancement of the symmetry group of the hydrogen atom to claim that we have a hidden $SU(d)$ symmetry in the problem. The mode function equation turns out to be essentially the Schr\"odinger equation for a quantum particle on a sphere with a spherical analog of the harmonic oscillator potential. This problem has been previously treated in \cite{Higgs,Leemon} (see also \cite{lakshesw}), and is known to possess a hidden $SU(d)$ symmetry. We will conclude by making some comments about how this symmetry might be responsible for the selection rules noticed in \cite{CEV1, BKS1, CEV2, Yang2}.

\section{Perturbation Theory, Resonances and Selection Rules} \label{selection}

We will start with a self-interacting scalar field in AdS$_{d+1}$, 
\begin{align}
S = \int d^{d+1}x \sqrt{-g} \, \left(\dfrac{1}{2}g^{\mu\nu}\partial_{\mu}\phi \ \partial_{\nu}\phi + V(\phi)\right),
\end{align}
with
\begin{align}
V(\phi) =\dfrac{m^2}{2}\phi^2+ \dfrac{\lambda}{(N+1)!} \phi^{N+1}.
\end{align}
The selection rule problem that drives our interest is only meaningful for some discrete values of the mass, most prominently including $m=0$, the specific value on which the previous investigations focused \cite{BKS1,Yang2,BKS2}. However, the hidden symmetries we want to display are present for any values of the mass, hence we shall for now treat it as arbitrary. The application of these hidden symmetries to selection rules should of course be discussed in the context of appropriate mass values.

The global AdS metric (after setting the AdS radius to unity) is
\begin{align}
ds^{2}_{AdS_{d+1}} = \sec^{2}x\left(-dt^{2} + dx^{2} + \sin^{2}x\, d\Omega_{d-1}^{2}\right)\equiv \sec^2 x \, (ds^2_{ES}), 
\label{AdSmetric}
\end{align}
where we have identifed the metric on the Einstein static universe $ds^2_{ES}$ for future convenience. (More specifically, only half of each spherical spatial slice of the Einstein static universe is included, since $x$ varies between $0$ and $\pi/2$, rather than $0$ and $\pi$. The resulting boundary is just a conformal image of the boundary of the AdS.)
We will use $\Omega$ to collectively denote the $(d-1)$-sphere coordinates appearing in $d\Omega_{d-1}$. The equations of motion for the scalar field are given by
\begin{align}\label{scalareom}
\Box_{AdS_{d+1}}\phi-m^2\phi \equiv \cos^{2}x\ \left(
-\partial_t^2\phi + \Delta^{(d)}_{s}\phi \right)
-m^2 \phi  = \dfrac{\lambda}{N!} \phi^{N}
\end{align}
where 
\bea 
\Delta^{(d)}_{s} \equiv  \frac{1}{\tan^{d-1} x} \partial_x(\tan^{d-1} x \partial_x)+\frac{1}{\sin^2x} \Delta_{\Omega_{d-1}}.
\eea
Here $\Delta_{\Omega_{d-1}}$ is the Laplacian on the  $(d-1)$-sphere.
The solution to the free theory (i.\ e., $\lambda=0$ in (\ref{scalareom})), which we shall denote $\phi^{(0)}$, can be found by separating variables, as presented (for example) in \cite{Hamilton}:
\bea
\label{GlobalModes}
\phi^{(0)}(t,x,\Omega) = \sum_{n = 0}^\infty \sum_{l,k} (A_{nlk}\ e^{-i\omega_{nlk}t}+\bar A_{nlk}\ e^{i\omega_{nlk}t}) e_{nlk}(x, \Omega),
\eea
where $A_{nlk}$ are arbitrary complex amplitudes and
\bea
\omega_{nlk}=2n + l + \Delta, \label{omega}
\eea
with
$\Delta= \frac{d}{2} + \sqrt{{d^2 \over 4} + m^2}$. The mode functions are
\bea
e_{nlk}(x, \Omega)=\cos^\Delta \! x \ \sin^l \! x\ P_n^{\left(\Delta - \frac{d}{2},\,l + \frac{d}{2} - 1\right)}
(-\cos 2 x) \ Y_{lk}(\Omega).
\label{modefunctions}
\eea
$Y_{lk}$ are the $(d-1)$-dimensional spherical harmonics, and the set of `azimuthal numbers' on a general $(d-1)$-sphere is collectively indicated by the label $k$. Its details will not affect the discussion below. $P^{(a,b)}_n(y)$ are the Jacobi polynomials. The mode functions satisfy the equation
\bea
\left(\Delta^{(d)}_{s}-\frac{m^2}{\cos^2 x}\right)e_{nlk}(x, \Omega)=-\omega_{nlk}^2 \ e_{nlk}(x, \Omega),
\label{adsmode}
\eea
and they are orthogonal with respect to a scalar product defined by
\beq
(f,g)=\int dx d\Omega\, \tan^{d-1}x\,f(x,\Omega)\, g(x,\Omega).
\label{scalarprod}
\eeq

One can then take the non-linearities into account perturbatively by expanding solutions to (\ref{scalareom}) as
\bea
\phi=\phi^{(0)}+\lambda\phi^{(1)}+\cdots
\eea
For $\phi^{(1)}$, one gets
\begin{align}\label{phi1}
-\partial_t^2\phi^{(1)} + \left(\Delta^{(d)}_{s}-\dfrac{m^2}{\cos^{2}x} \right)\phi^{(1)} 
 = \dfrac{\lambda}{N!} \dfrac{(\phi^{(0)})^{N}}{\cos^{2}x}.
\end{align}
It is convenient to expand $\phi^{(1)}$ in the basis of $e_{nlk}$:
\bea
\label{phi1Modes}
\phi^{(1)}(t,x,\Omega) = \sum_{nlk} c^{(1)}_{nlk}(t) e_{nlk}(x, \Omega).
\eea
Substituting (\ref{GlobalModes}), (\ref{adsmode}) and (\ref{phi1Modes}) in (\ref{phi1}), and projecting on the eigenmodes $e_{nlk}$ using (\ref{scalarprod}) one gets
\begin{align}\label{c1}
&\ddot c^{(1)}_{nlk} +\omega^2_{nlk} c^{(1)}_{nlk}\sim \dfrac{\lambda}{N!}\sum\limits_{n_1l_1k_1}\cdots\sum\limits_{n_Nl_Nk_N} C_{nlk|n_1l_1k_1|\cdots|n_{N}l_{N}k_{N}}\\
&\nonumber(A^{(0)}_{n_1l_1k_1}\ e^{-i\omega_{n_1l_1k_1}t}+\bar A^{(0)}_{n_1l_1k_1}\ e^{i\omega_{n_1l_1k_1}t})\cdots(A^{(0)}_{n_{N}l_{N}k_{N}}\ e^{-i\omega_{n_{N}l_{N}k_{N}}t}+\bar A^{(0)}_{n_{N}l_{N}k_{N}}\ e^{i\omega_{n_{N}l_{N}k_{N}}t}),
\end{align}
where we have used a proportionality (rather than equality) sign since we have not kept track of the normalization of the mode functions, which is inessential for our purposes. The coefficients $C$ are given by
\bea
C_{nlk|n_1l_1k_1|\cdots|n_{N}l_{N}k_{N}}=\int dx d\Omega\, \tan^{d-1}x \sec^2x\,\,e_{nlk} e_{n_1l_1k_1}\cdots e_{n_Nl_Nk_N}.
\label{Ccoeff}
\eea

The right-hand side of (\ref{c1}) consists of a sum of simple oscillating terms of the form $e^{i\omega t}$ with 
\beq
\omega=\pm\omega_{n_1l_1k_1}\pm\cdots\pm \omega_{n_Nl_Nk_N},
\eeq
where all the plus-minus sign choices are independent. If $\omega$ is different from $\pm\omega_{nlk}$, the corresponding term produces an innocuous oscillating contribution to $c^{(1)}$, and the corresponding contribution to $\phi$ remains bounded, with an amplitude proportional to $\lambda$ for all times. However, if $\omega=\pm\omega_{nlk}$, the corresponding term is in {\em resonance} with the left-hand side of (\ref{c1}), producing a {\em secular} term in $c^{(1)}$ and $\phi$, a term that will grow indefinitely with time and invalidate the perturbation theory at times of order $1/\lambda$.

In order to make the perturbation theory usable for large times (which is the regime of physical interest), different resummation schemes can be employed, such as the two-time formalism \cite{Buchel1, BKS1, BKS2}, renormalization group resummation \cite{CEV1,CEV2} and averaging \cite{BKS1, CEV2}. Some systematic discussion of resummation techniques in the context of AdS dynamics can be found in  \cite{CEV1,CEV2}. All the resummation schemes mentioned are equivalent at the lowest non-trivial order, which is our present setting. The result of resummation procedures is an improved perturbation theory valid at times of order $1/\lambda$ in which secular terms become replaced with (resummed) slow changes of the mode amplitudes. The picture of non-linearities inducing energy transfer between linear modes (this transfer being slow in the weakly non-linear regime) is physically very intuitive.

Each resonant term in the sum on the right hand side of (\ref{c1}) produces the corresponding term in $c_1$ and, after resummation, a corresponding term in the flow equations describing the slow variations of the complex amplitudes, cf. (\ref{GlobalModes}), due to the energy transfer between the modes. The general resonance condition reads
\beq
\omega_{nlk}=\pm\omega_{n_1l_1k_1}\pm\cdots\pm \omega_{n_Nl_Nk_N}.
\label{omegares}
\eeq
It has been noted, however, that in AdS settings, many of the plus-minus choices in the above expression do not in fact result in secular terms, because the corresponding $C$-coefficients in (\ref{c1}) vanish. In \cite{CEV1}, this phenomenon was proved for spherically symmetric perturbations of dynamical gravity coupled to a scalar field. In \cite{BKS1}, it was noted that similar vanishing occurs for spherically symmetric configurations of a self-interacting scalar field in a fixed AdS geometry, which is our present setting. In \cite{Yang2}, the assumption of spherical symmetry was relaxed and a powerful set of {\em selection rules} was given for the $C$-coefficients as defined by (\ref{Ccoeff}).

More specifically, the considerations of \cite{Yang2} established that, if $d(N+1)$ is even and $m=0$, the resonances corresponding to choosing all plus signs in (\ref{omegares}) always drop out from the dynamics due to the vanishing of the corresponding $C$-coefficients. Using (\ref{omega}), this particular resonance condition can be re-written as
\beq
2n+l=(N-1)\Delta+2n_1+l_1+\cdots+2n_N+l_N.
\eeq
Note that the resonance condition itself cannot be satisfied for general values of the mass $m$, since $\Delta$ is in general non-integer. However, $m=0$ implies $\Delta=d$, in which case there are always some modes satisfying the above condition. Despite the fact that the resonance condition is satisfied, however, the vanishing of the corresponding $C$-coefficients results in the absence of the corresponding secular terms and energy transfer channels. This, in turn, finds expression in extra conservation laws restricting the slow energy transfer, analyzed in \cite{BKS1,CEV2,Buchel2,Yang2}.

\section{Hidden $SU(d)$ Symmetry and Mode Function Multiplets}\label{wave}

In the previous section, we have reviewed non-linear perturbation theory for scalar fields in AdS space-time and the emergence of resonances generating significant (slow) energy transfer between the modes for arbitrarily small non-linearities. We have displayed a set of {\em selection rules} forcing some of these resonances to vanish despite they could be present on general grounds. In all the previous analytic considerations \cite{CEV1,Yang2}, the selection rules were proved using brute force manipulations involving the properties of orthogonal polynomials contained in (\ref{modefunctions}). It is natural to believe that there exists a more qualitative explanation for the selection rules, most likely based on the high degree of symmetry and other special properties of the underlying AdS background. The concept of the ground state symmetries restricting the energy flow between the perturbation modes due to non-linearities is familiar in more conventional settings, such as vibrations of crystalline lattices \cite{chechin}. Similar suspicions of the symmetry origins of the AdS selection rules have been voiced in \cite{Yang2}.

One is thus confronted with the question of the symmetry properties of the mode functions (\ref{modefunctions}) appearing in the integrals (\ref{Ccoeff}). Here, one immediately observes an intriguing structure. The mode functions are eigenfunctions defined by (\ref{adsmode}), and each set of eigenfunctions with the same eigenvalue $\omega^2_{nlk}$ must form an irreducible representation of the symmetry group of the operator on the left-hand side of (\ref{adsmode}), whose eigenvalue problem is studied. The only obvious symmetry this operator ($\Delta^{(d)}_{s}-{m^2}/{\cos^2 x}$) has is the $SO(d)$ rotations of the $(d-1)$-sphere parametrized by $\Omega$. This symmetry explains why the eigenvalues (\ref{omega}) do not depend on the `azimuthal numbers' $k$. The degeneracy is much higher however, since the eigenvalues are not only independent of $k$, but also only depend on $l$ and $n$ through the combination $2n+l$. Different representations of the rotational $SO(d)$, labeled by $l$, are bundled together to form much bigger representations, of what must be a bigger symmetry group. What can this group be?

One might have been tempted to look for isometries of AdS as the source of degeneracies. Since the mode functions are defined on a single spatial slice, rather than in the whole space-time, one might have tried to talk of the isometry group of a single spatial slice of AdS$_{d+1}$, which is $SO(d,1)$. This is a wrong perspective, however, since the Laplacian on a single spatial slice of AdS$_{d+1}$ is $\cos^2 x \,\Delta^{(d)}_{s}$, which is different from the operator on the left-hand side of equation (\ref{adsmode}) defining the mode functions, even for $m=0$. Not surprisingly, representations of $SO(d,1)$ do not decompose into representations of its rotational subgroup $SO(d)$ in a way compatible with the AdS frequency degeneracies. The eigenvalues in (\ref{adsmode}) depend only on $2n+l$, which implies that the values of the angular momentum $l$ entering each `level' are either all even or all odd. This property is not shared by the $SO(d)$ decomposition of $SO(d,1)$ representations.

To reveal the actual enhanced symmetry group of (\ref{adsmode}) it is convenient to first recall the conformal relation (\ref{AdSmetric}) between the AdS metric and the Einstein static metric. Using the standard conformal transformation formulas, displayed, for instance, in (3.5) of \cite{BirrellDavies}, one can obtain the following identity for arbitrary $\phi(t,x,\Omega)$:
\bea\label{AdSES}
\cos^{(d+3)/2}x\left(\Box_{ES}-\frac{(d-1)^2}{4}\right) \frac{\phi(t,x,\Omega)}{\cos^{(d-1)/2} x} =
\left(\Box_{AdS_{d+1}}+\frac{d^2-1}{4}\right)\phi (t,x,\Omega).\hspace{4mm}
\eea
Let us emphasize for clarity that on the left hand side the operator is acting on $\phi(t,x,\Omega)/\cos^{(d-1)/2} x$. Now, 
\bea
\Box_{ES}=-\partial_t^2+\Delta_{\Omega_d},
\eea
where the $d$-sphere Laplacian is explicitly
\bea 
\Delta_{\Omega_d} \equiv  \frac{1}{\sin^{d-1} x} \partial_x(\sin^{d-1} x \ \partial_x)+\frac{1}{\sin^2x} \Delta_{\Omega_{d-1}}. 
\label{dsphL}
\eea
Correspondingly, by relating the AdS wave function equation to the ES wave equation using (\ref{AdSES}), one can re-write the mode function equation (\ref{adsmode}) in the form of a Schr\"odinger equation
\bea
\left(-\Delta_{\Omega_d}+V(x)\right) \tilde e_{nlk}=E_{nlk} \tilde e_{nlk}, \label{schrod}
\eea
with
\bea
V(x)=\frac1{\cos^2 x}\left(m^2+\frac{d^2-1}{4}\right)\qquad\mbox{and}\qquad E_{nlk}=\omega_{nlk}^2-\frac{(d-1)^2}4.
\eea
We have defined $\tilde e_{nlk} \equiv e_{nlk}/\cos^{(d-1)/2} x$ for convenience. Note that the range of $x$ is only a hemi-sphere, $x \in [0, \pi/2)$, since the potential $V(x)$ is unbounded and confines the `particle' to this hemi-sphere.

Enhanced symmetries of the Schr\"odinger equation for a particle on a $d$-sphere, with a potential $V(x) \sim 1/\cos^2 x$ have been studied in \cite{Higgs, Leemon} from a purely quantum-mechanical perspective. (It may also be useful to consult \cite{KMP}, which uses notation more similar to ours.) Historically, the equations were solved in \cite{lakshesw} and the observed abnormal energy degeneracies of the sort we described above (with energies depending only on $2n+l$) prompted an investigation into enhanced symmetries. 

The easiest way to notice the presence of enhanced symmetries in (\ref{schrod}) is by looking at the corresponding classical problem. Solving the equations of motion in centrally symmetric potentials is standard and we shall not review the details here. It is easy to show that if the orbital shape of a trajectory in a central potential $V(r)$ in the ordinary $d$-dimensional flat space is $r=r(\varphi)$, then the orbital shape of the motion on a $d$-sphere in the potential $V(\tan x)$ is $x=\arctan(r(\varphi))$, with $x$ being the polar angle on the sphere, as in (\ref{schrod}). It is a simple corollary that if the orbits close for a central potential $V(r)$ in the ordinary $d$-dimensional flat space, then they will close as well for the central potential $V(\tan x)$ on a $d$-sphere.

It is known that the orbits close in flat space only for two potentials: the Coulomb potential $V(r)\sim 1/r$ and the isotropic harmonic oscillator potential $V(r)\sim r^2$. This is the so-called Bertrand's theorem \cite{Bertrand}. The corresponding potentials on a $d$-sphere, for which the closure of orbits is guaranteed by the above consideration, are the sphere Coulomb potential $V(x)\sim \cot x$ and the sphere `harmonic oscillator' potential $V(x)\sim 1/\cos^2 x$. It is the latter potential that appears in the `Schr\"odinger' equation (\ref{schrod}).

In flat space, the closure of orbits is explained by enhanced symmetries and the corresponding conserved quantities. For the Coulomb potential $V(r)\sim 1/r$, the conserved quantity is the Laplace-Runge-Lenz vector, which, together with the angular momentum, forms an $so(d+1)$ Lie algebra with respect to taking the Poisson brackets. For the isotropic harmonic oscillator potential $V(r)\sim r^2$ the corresponding conserved quantity is a traceless symmetric second rank tensor, known as Elliott's quadrupole after \cite{elliott}, or the Fradkin tensor after \cite{fradkin}. Together with the angular momentum (antisymmetric second rank tensor), it forms an $su(d)$ Lie algebra.

The situation on a $d$-sphere forms a close parallel to the one described above for flat space. The d-sphere Coulomb problem first appeared in \cite{schr} and reveals an $SO(d+1)$ symmetry and the corresponding degeneracy pattern, exactly identical to the flat space case. The sphere harmonic oscillator, which is our main interest here, has been analyzed in \cite{Higgs, Leemon}. The classical version of the corresponding $SU(d)$ symmetry has been displayed and the corresponding conserved quantities have been constructed. A quantum version of this symmetry, leaving equation (\ref{schrod}) invariant, has only been constructed \cite{Higgs} for $d=2$, which corresponds to $AdS_3$ in our setting, because of the problems with resolving the ordering ambiguities. Its explicit construction remains an outstanding technical problem, to the best of our knowledge. Nevertheless, the fact that the degeneracies of the energy levels of (\ref{schrod}) and their properties under $SO(d)$ spatial rotations fit representations of $SU(d)$ demonstrates that the classical $SU(d)$ symmetry is in no way upset by quantization.

We would like to emphasize that the symmetry, the degeneracies and the multiplets furnished by the eigenfunctions are exactly the same for our sphere harmonic oscillator as for the usual straightforward isotropic harmonic oscillator in flat space. Of course, the symmetries are realized in a much more non-trivial way in the non-linear case of motion on a sphere. For flat space, the $SU(d)$ symmetry can be seen immediately by simply writing the Hamiltonian $H\sim\sum(p_i^2+x_i^2)$ in terms of the creation-annihilation operators $H\sim\sum a^\dagger_i a_i$. Transforming $a_i$ to $\tilde a_i=S_{ik}a_k$, with any $SU(d)$ matrix $S_{ik}$, obviously leaves the Hamiltonian invariant. It is furthermore straightforward, due to the linear nature of the flat space harmonic oscillator, to implement any such transformation as a unitary operator on the Hilbert space, $\tilde a_i=U a_i U^\dagger$.

We finally identify the $SU(d)$ transformation properties of the multiplets corresponding to each `energy' level in (\ref{schrod}). These can be easily reconstructed from the transformation properties of the mode functions under the obvious $SO(d)$ subgroup of $SU(d)$ representing the spatial rotations in (\ref{schrod}). For each given $n$ and $l$, the rotational properties of the mode functions are determined by the spherical harmonics $Y_{lk}(\Omega)$, which transform according to the traceless symmetric rank $l$ tensor representation of $SO(d)$. A given representation of $SU(d)$ is formed by all such functions with the same value of `energy', i.\ e., with the same value of $2n+l$. Since both $n$ and $l$ are positive, there will be a maximal possible value of $l$ in each multiplet, which we shall call $L$. Each level will then be composed of the following $SO(d)$ multiplets: $(n=0,l=L)$, $(n=1,l=L-2)$, etc. This is precisely the $SO(d)$ content of the fully symmetrized $L$th power of the fundamental representation of $SU(d)$. Indeed, to separate irreducible representations of $SU(d)$ into irreducible representations of its $SO(d)$ subgroup, one must separate each tensor into its trace and traceless parts \cite{hamermesh}. Applied to a fully symmetric tensor of rank $L$, this will generate traceless fully symmetric tensors of ranks $L$, $L-2$, etc (since two indices get contracted to produce each trace). These are precisely the rotational representations appearing for mode functions at each given level, with the angular momentum varying in steps of 2.

\section{Comments on the Symmetry Origin of the Selection Rules}

Having established that the mode functions (\ref{modefunctions}) of frequency $\Delta+L$ form multiplets transforming as the fully symmetrized $L$th power of the fundamental representation of $SU(d)$, the hidden symmetry group of the mode function equation (\ref{adsmode}), one might wonder what repercussions this observation has on the selection rules for the energy flow coefficients (\ref{Ccoeff}).

One may first recall how selection rules arise in more conventional settings, for example, for cases with ordinary spherical symmetry $SO(d)$. Consider an integral of spherical harmonics
\beq
\int d\Omega\,Y_{l_1k_1}(\Omega)\cdots Y_{l_Nk_N}(\Omega).
\label{rotation}
\eeq
One can use the standard angular momentum addition theory to decompose the product of spherical harmonics into a sum of irreducible representations of $SO(d)$. Integrated over all angles, any non-trivial (non-scalar) irreducible representation will produce a zero result. The only way the integral can be non-zero is if the trivial (scalar) representation is contained in the direct product of the representations corresponding to $Y_{l_1k_1},\cdots, Y_{l_Nk_N}$. By the usual addition of angular momenta, this can only happen if each $l_i$ is less than or equal to the sum of all the other $l_i$. Hence the angular momentum selection rules.

The application of $SU(d)$ symmetry to the integral (\ref{Ccoeff}) is considerably less straightforward for the following reasons. First, the $SU(d)$ is not made of purely spatial transformations. In the Schr\"odinger equation language of (\ref{schrod}) it is a quantum symmetry originating from classical canonical transformations mixing coordinates and momenta.\footnote{The $SU(d)$ hidden symmetry group we find is closely related to the isometry group of the complex projective space $\C\PP^{d-1}$. Indeed, the corresponding geometric structure can be revealed in the phase space of the ordinary $d$-dimensional flat space  isotropic harmonic oscillator, as in section 5.4.5.3 of \cite{geodyn}. If an analogous representation is found for the $d$-sphere case, it may turn out useful for our pursuits.} It therefore does not act on the integral (\ref{Ccoeff}) as straightforwardly as spatial rotations act on (\ref{rotation}). Even more frustratingly, explicit construction of the symmetry generators for (\ref{schrod}) has evaded dedicated effort in \cite{Higgs,Leemon}, except for the relatively simple case $d=2$. This is despite the fact that the symmetry is certainly there, as evidenced by the symmetries of the corresponding classical problem, and the level degeneracies and wave-function rotational multiplets of the quantum problem matching the decomposition of multiplets of the $SU(d)$ in terms of its rotational subgroup $SO(d)$. It is these technical complications that encouraged us to present our understanding in its current form, postponing more detailed investigations to future work.

Despite the above technical complications, one might envisage the possible algebraic patterns responsible for the selection rules in (\ref{Ccoeff}) in the presence of the $SU(d)$ symmetry. The symmetry transformations connect the mode functions $e_{nlk}$ in the same multiplet, i.\ e., with the same values of $2n+l$. There must exist the corresponding conjugate raising and lowering operators $\hat A_{\pm}$, increasing (decreasing) $l$ by 2 and decreasing (increasing) $n$ by 1. Analogous operators have been constructed for the simple flat space isotropic harmonic oscillator in \cite{liuleizeng}. Imagine then that $2n+l=L$ in (\ref{Ccoeff}) is large. One can write $e_{nlk}$ as a certain number of lowering operators $\hat A_{-}$ acting on a mode function in the same multiplet with the highest value of $l$, i.\ e., $e_{0L\tilde k}$. It should be possible to use the conjugation properties of  $\hat A_{\pm}$ to turn the $\hat A_{-}$ acting on $e_{0L\tilde k}$ into  $\hat A_{+}$ acting on the remaining mode functions under the integral, which will raise their angular momentum values somewhat. The result will be an integral of a product of $e_{0L\tilde k}$, carrying a very high angular momentum $L$, with transformed  $e_{n_il_ik_i}$. Note that acting with $SU(d)$ symmetry transformations on  $e_{n_il_ik_i}$ can never increase their angular momentum beyond $2n_i+l_i$, because there are no such states in the multiplets. In the end, after applying the $SU(d)$ transformation to the integral (\ref{Ccoeff}) one will end up with an integral of a product of a mode function with a very large angular momentum $2n+l$ and other mode functions with angular momenta of at most $2n_i+l_i$. This product is strongly constrained by the {\em ordinary} angular momentum selection rules of the sort described under (\ref{rotation}). Note that, after the $SU(d)$ transformation has been applied, restrictive angular momentum constraints will arise even if all the values of $l$ and $l_i$ in the original integral (\ref{Ccoeff}) were zero, in which case a direct application of angular momentum selection rules would have been completely vacuous.

At a practical level, there is a number of technical obstructions to implementing the above program in a detailed fashion. As we have already mentioned, we are not aware of an explicit construction of the $SU(d)$ symmetry generators for (\ref{schrod}), and the attempts of \cite{Higgs, Leemon} have been plagued by algebraic difficulties. This precludes a straightforward specification of the raising and lowering operators, unlike the much more obvious flat space case of \cite{liuleizeng}. One has to worry, furthermore, about the trigonometric insertions in (\ref{Ccoeff}). Part of those insertions will be absorbed by the integration measure necessary to make the raising and lowering operators conjugate. What remains will be acted upon by the raising operators, if one attempts the construction in the previous paragraph, and its $SU(d)$ transformation properties will have to be discussed explicitly before the detailed form of the resulting selection rules can be exposed. All of this would require a more direct understanding of how the $SU(d)$ symmetry acts on the Hilbert space of (\ref{schrod}).

\section{Acknowledgments}

We would like to thank George Chechin for a useful exposition on the role of symmetry constraints in non-linear vibrations of atomic lattices, Martin Cederwall for a very stimulating discussion on AdS isometries and their representations, Andrzej Rostworowski for useful correspondence, and Pallab Basu, Ben Craps, P. N. Bala Subramanian and Joris Vanhoof for collaboration on closely related subjects. The work of O.~E. has been supported by the Ratchadaphisek Sompote Endowment Fund.


\begin{thebibliography}{20}
\bibitem{BizonReview} P. Bizo\'n, \emph{Is AdS stable?}, Gen.\ Rel.\ Grav.  {\bf 46} (2014) 1724, [arXiv:1312.5544].
\bibitem{Bizon1} P. Bizo\'n and A. Rostworowski, \emph{On weakly turbulent instability of anti-de Sitter
space}, Phys.\ Rev.\ Lett. {\bf 107} (2011) 031102, [arXiv:1104.3702]. 
\bibitem{Maliborski:2013jca}
  M.~Maliborski and A.~Rostworowski,
  \emph{Time-Periodic Solutions in an Einstein AdS-Massless-Scalar-Field System},
  Phys.\ Rev.\ Lett.\  {\bf 111} (2013) 051102,
 [arXiv:1303.3186].

\bibitem{Buchel:2013uba}
  A.~Buchel, S.~L.~Liebling and L.~Lehner,
  \emph{Boson stars in AdS spacetime},\\
  Phys.\ Rev.\ D {\bf 87} (2013) 123006,
  [arXiv:1304.4166].
\bibitem{Buchel1} V. Balasubramanian, A. Buchel, S. R. Green, L. Lehner, S. L. Liebling, \emph{Holographic Thermalization, stability of AdS, and the Fermi-Pasta-Ulam-Tsingou paradox}, Phys. Rev. Lett. {\bf 113} (2014) 071601, [arXiv:1403.6471]
\bibitem{Bizon2} P. Bizo\'n and A. Rostworowski, \emph{Comment on "Holographic Thermalization, stability of AdS, and the Fermi-Pasta-Ulam-Tsingou paradox" by V. Balasubramanian et al}, [arXiv:1410.2631].
\bibitem{CEV1} B. Craps, O. Evnin and J. Vanhoof, \emph{Renormalization group, secular term resummation and AdS (in)stability}, JHEP {\bf 10} (2014) 48, [arXiv:1407.6273].
\bibitem{BKS1} P. Basu, C. Krishnan and A. Saurabh, \emph{A Stochasticity Threshold in Holography and
the Instability of AdS}, [arXiv:1408.0624].
\bibitem{Yang1} F. V. Dimitrakopoulos, B. Freivogel, M. Lippert, I-S. Yang, \emph{Instability corners in AdS space}, [arXiv:1410.1880].
\bibitem{CEV2} B. Craps, O. Evnin and J. Vanhoof, \emph{Renormalization, averaging, conservation laws and AdS (in)stability}, JHEP {\bf 1501} (2015) 108, [arXiv:1412.3249].
\bibitem{Buchel2} A. Buchel, S. R. Green, L. Lehner, S. L. Liebling, \emph{Conserved quantities and dual turbulent cascades in Anti-de Sitter spacetime}, [arXiv:1412.4761].
\bibitem{Yang2} I-S. Yang, \emph{The missing top of AdS resonance structure}, [arXiv:1501.00998]
\bibitem{BKS2} 
  P.~Basu, C.~Krishnan and P.~N.~B.~Subramanian,
  \emph{AdS (In)stability: Lessons From The Scalar Field},
  [arXiv:1501.07499].

\bibitem{chechin} G.~M.~Chechin, V.~P.~Sakhnenko, \emph{Interactions between normal modes in nonlinear dynamical systems with discrete symmetry. Exact results}, Physica\ D {\bf 117} (1998) 43.

\bibitem{Higgs} 
  P.~W.~Higgs,
  \emph{Dynamical Symmetries in a Spherical Geometry 1},\\
  J.\ Phys.\ A {\bf 12}  (1979) 309.

\bibitem{Leemon} 
  H.~I.~Leemon,
  \emph{Dynamical Symmetries in a Spherical Geometry 2},\\
  J.\ Phys.\ A {\bf 12}  (1979) 489.


\bibitem{lakshesw}
M.~Lakshmanan and K.~Eswaran, \emph{Quantum dynamics of a solvable nonlinear chiral model},  J. Phys. A {\bf 8} (1975) 1658.

\bibitem{Hamilton} 
  A.~Hamilton, D.~N.~Kabat, G.~Lifschytz and D.~A.~Lowe,
  \emph{Holographic representation of local bulk operators},
  Phys.\ Rev.\ D {\bf 74}  (2006) 066009,
  [hep-th/0606141].

\bibitem{BirrellDavies}
N.~D.~Birrell, P.~C.~W.~Davies, \emph{Quantum Fields in Curved Space}, CUP (1982).

\bibitem{KMP}
E.~G.~Kalnins, W.~Miller,  G.~S.~Pogosyan, \emph{The Coulomb-Oscillator Relation 
on n-Dimensional Spheres and Hyperboloids}, Phys. Atom. Nucl. {\bf 65} (2002) 1086,
 [math-ph/0210002].

\bibitem{Bertrand} J.~Bertrand, \emph{Th\'eor\`eme relatif au mouvement d'un point attir\'e vers un centre fixe},\\ C. R. Acad. Sci. {\bf 77} (1873) 849.

\bibitem{elliott}
J.~P.~Elliott, \emph{Collective Motion in the Nuclear Shell Model I.
Classification Schemes for States of Mixed Configurations}, Proc.\ R.\ Soc.\ Lond.\ A {\bf 245} (1958) 128.

\bibitem{fradkin}
D.~M.~Fradkin,  \emph{Three-Dimensional Isotropic Harmonic Oscillator and $SU_3$},\\
 Am.\ J.\ Phys. {\bf 33} (1965) 207.

\bibitem{schr}
E.~Schr\"odinger, \emph{A Method of Determining Quantum-Mechanical Eigenvalues and Eigenfunctions}, Proc. Royal Irish Acad. A {\bf 46} (1940/1941) 9.

\bibitem{hamermesh}
M.~Hamermesh, \emph{Group Theory and Its Application to Physical Problems}, New York: Dover (1989). 

\bibitem{geodyn}
J.~F.~Cari\~{n}ena, A.~Ibort, G.~Marmo, G.~Morandi, \emph{Geometry from Dynamics, Classical and Quantum}, Springer (2014).

\bibitem{liuleizeng}
Y.~F.~Liu, Y.~A.~Lei, J.~Y.~Zeng, \emph{Factorization of the radial Schr\"odinger equation and four kinds of raising and lowering operators of hydrogen atoms and isotropic harmonic oscillators}, Phys. Lett. A {\bf 231} (1997) 9.

\end{thebibliography}
\end{document}